# Augmented Reality Indoor Wayfinding in Hospital Environments

*An Empirical Study on Navigation Efficiency, User Experience, and Cognitive Load*

*By*

Kai Liu[1], Michelle L. Aebersold[2], Mark Lindquist[3], Haoting Gao[4]

[1] University of Michigan School of Information
[2] University of Michigan School of Nursing
[3] University of Michigan School for Environment and Sustainability
[4] University of Michigan Medical School

## ABSTRACT

Hospitals are among the most cognitively demanding indoor environments, especially for patients and visitors unfamiliar with their layout. This study investigates the effectiveness of an augmented reality (AR)–based handheld navigation system compared to traditional paper maps in a large hospital setting. Through a mixed-methods experiment with 32 participants, we measured navigation performance, cognitive workload (NASA-TLX), situational anxiety (STAI-State), spatial behavior, and user satisfaction. Results show that AR users completed navigation tasks significantly faster, made fewer errors, and reported lower anxiety and workload. However, paper map users demonstrated stronger spatial memory in sketch-based recall tasks, highlighting a trade-off between real-time efficiency and long-term spatial learning. We discuss implications for inclusive AR design, spatial cognition, and healthcare accessibility, offering actionable design strategies for adaptive indoor navigation tools.

## Keywords

## 1 INTRODUCTION

Hospitals are among the most spatially complex built environments, featuring multi-level layouts, intersecting corridors, and a wide range of specialized departments. For patients and visitors who are unfamiliar with the space, navigating such facilities can be both confusing and stressful—frequently leading to delays, disorientation, and heightened anxiety (Abu-Obeid, 1998; Ulrich et al., 2010). These wayfinding challenges also have operational consequences, as they often disrupt clinical workflows by diverting staff attention away from patient care (Lee et al., 2021).

The concept of *wayfinding*, introduced by Kevin Lynch in *The Image of the City*, refers to the process by which individuals use environmental cues—such as paths, landmarks, nodes, and

districts—to build a mental representation of a space. Hospitals, much like small cities, contain complex circulation systems, departmental clusters, and functional zones, which often compromise spatial legibility for first-time users (Abu-Ghazzeh, 1996).

Unlike outdoor navigation systems that leverage GPS, indoor settings such as hospitals pose distinct technological constraints. Poor signal reception, dense architectural structures, and a lack of dynamic guidance hinder the effectiveness of traditional navigation tools (Morag & Pintelon, 2021; Sánchez-Jáuregui Descalzo et al., 2024). While mobile apps and digital patient portals have become more common, these tools often fall short in supporting spatial orientation and reducing cognitive effort during indoor navigation (Zhou et al., 2022). This limitation highlights a growing need for context-aware, user-centered solutions specifically designed to enhance wayfinding in healthcare environments.

This study explores augmented reality (AR)-based indoor navigation as a promising alternative to traditional wayfinding methods. By overlaying visual cues—such as arrows, labels, and directional prompts—onto the user's view of the physical environment, AR offers spatially contextualized guidance that may reduce both mental workload and navigation-related stress. As AR technologies become increasingly accessible through smartphones and tablets, their potential for deployment in public healthcare settings warrants deeper empirical investigation.

To address this opportunity, we conducted a real-world study at the University of Michigan Hospital, evaluating the effectiveness of an AR-based navigation system compared to conventional paper maps. This research aims to:

- Assess whether AR navigation reduces cognitive load and spatial anxiety during hospital wayfinding tasks;
- Examine the impact of AR-guided navigation on user experience, task efficiency, and spatial orientation in complex indoor environments;
- Derive actionable design recommendations for future AR-based wayfinding systems in healthcare settings.

Through a mixed-methods, user-centered evaluation, this work contributes to ongoing discussions in HCI, spatial cognition, and healthcare design, offering both theoretical insights and practical implications for improving patient experience and operational efficiency in large-scale medical facilities.

## 2 COMPLEX BUILT ENVIRONMENTS AND WAYFINDING CHALLENGES IN HOSPITALS

### 2.1 Physical Complexity: Multi-Level Layouts and Signage Limitations

Hospitals are architecturally complex, often comprising interconnected buildings, vertical and horizontal circulation networks, and departments spread across multiple floors. Such spatial dispersion frequently overwhelms first-time visitors, who lack intuitive directional cues. This complexity complicates the creation of universally effective signage, especially given the diverse spatial needs of users (Gutiérrez Pérez et al., 2024; Rooke et al., 2010).

### 2.2 Dynamic Conditions: High Traffic and Constant Change

Hospital environments are dynamic and often unpredictable. High foot traffic, frequent emergencies, construction zones, noise, and changing floor plans disrupt navigation and increase cognitive burden. These ever-changing conditions limit the reliability of static signage and printed maps, calling for adaptive, real-time navigation systems (Morag & Pintelon, 2021).

### 2.3 Cognitive and Emotional Demands: Anxiety and Overload

Patients and visitors often face heightened anxiety and cognitive overload, especially during initial hospital visits. Emotional distress impairs the ability to process complex signage and spatial cues (Carpman & Grant, 2016). Existing tools like patient portals often lack spatial support, offering little assistance in orientation (Zhou et al., 2022).

### 2.4 Systemic Limitations: Inflexible and Outdated Tools

Most hospitals still rely on static signs, printed directories, or basic 2D digital maps. These systems often fail to meet users' real-time needs, lacking interactivity, personalization, and adaptability to environmental changes. Effective wayfinding requires users to locate themselves, identify destinations, plan routes, and recognize arrival—tasks that current systems frequently fail to support (Huelat, n.d.; Sánchez-Jáuregui Descalzo et al., 2024).

## 3 IMPACTS AND BENEFITS OF EFFECTIVE WAYFINDING IN HOSPITALS

### 3.1 Consequences of Inadequate Wayfinding

#### (1) Negative Patient Experience and Anxiety

Wayfinding difficulties are a major non-clinical burden for hospital visitors, often resulting in heightened stress, confusion, and physiological symptoms such as fatigue and elevated blood pressure (Ulrich et al., 2010; Lee et al., 2020). When patients cannot find their destinations independently, it undermines their sense of control and contributes to a poorer overall care experience.

#### (2) Operational Disruptions and Institutional Costs

Poor navigation also affects hospital operations. Frequent direction requests divert staff from clinical duties and contribute to workflow interruptions (Carpman et al., 1990). Improved wayfinding has been associated with increased operational efficiency and measurable cost savings, such as reduced idle time among clinical staff (Lee et al., 2020).

#### (3) Exacerbated Impact in Resource-Constrained Settings

In overcrowded or under-resourced hospitals, unclear signage can cause significant delays and confusion, particularly for patients with disabilities or low literacy. In such contexts, wayfinding challenges can directly impact access to care and appointment adherence (Upadhyay et al., 2022).

### 3.2 Benefits of Effective Wayfinding Systems

#### (1) Improved Confidence and Reduced Stress

Effective navigation tools restore user autonomy and reduce emotional distress (Passini et al., 2000). However, traditional signage and 2D maps are often cluttered, outdated, or difficult to interpret, especially under time pressure or emotional strain.

**(2) Reduced Cognitive Load and Greater Accessibility**

Visual cues such as simplified pathways and color-coded markers help lower the cognitive burden of navigation, especially for users with dementia, ADHD, or other cognitive impairments (Passini et al., 2000). Intuitive systems also improve accessibility for a wider range of users.

**(3) Operational Gains and Cost Reduction**

Efficient wayfinding leads to shorter travel distances, fewer wrong turns, and fewer staff interruptions—producing measurable gains in staff productivity and reduced overhead costs (Lee et al., 2020).

**(4) Equity in High-Density, Low-Resource Settings**

Multimodal wayfinding systems—combining icons, voice prompts, and tactile guidance—can reduce dependence on staff while improving equity of care in complex or resource-constrained facilities (Upadhyay et al., 2022).

**3.3 Limitations of Current Navigation Systems**

**(1) Static Signage and 2D Maps**

Traditional maps and signs often provide excessive or unclear information, increasing cognitive demands. Additionally, they struggle to keep pace with frequent environmental changes (Colette, 2011).

**(2) Non-Interactive Digital Tools**

Most hospital apps and kiosks offer static 2D maps requiring users to mentally translate symbols into physical movements. This abstraction increases navigation errors and cognitive load (Dong et al., 2021).

**(3) Lack of Real-Time and Contextual Support**

Hospitals rarely use immersive or context-aware guidance tools. Existing systems often lack dynamic updates or adaptive feedback (Gardony et al., 2018).

**(4) Poor Accessibility and Personalization**

Many systems fail to address the needs of users with visual, cognitive, or linguistic challenges.

Despite the promise of AR and voice-guided systems, these are rarely implemented at scale (Chang, Tsai, & Wang, 2008).

### 3.4 Toward an AR-Based Navigation Approach

Given the limitations of static and non-interactive systems, hospitals require more adaptive and user-centered navigation technologies. The growing availability of AR-capable mobile devices offers a unique opportunity to reimagine hospital wayfinding through personalized, context-aware, and cognitively supportive solutions. The following chapter introduces an AR-based system designed to address these challenges through real-time spatial anchoring and intuitive guidance.

## 4 AR NAVIGATION AS A SOLUTION TO HOSPITAL WAYFINDING CHALLENGES

### 4.1 Overview of Indoor Navigation Technologies

Several technologies have been developed to support indoor navigation in complex environments, including Wi-Fi triangulation, Ultra-Wideband (UWB), Bluetooth beacons, LiDAR scanning, and vision-based Simultaneous Localization and Mapping (SLAM) (Zou et al., 2021; Sukhareva et al., 2021; Xu et al., 2024). While each has shown promise in controlled settings, their adoption in real-world healthcare environments remains limited (Romli et al., 2020; Tran & Parker, 2020).

Wi-Fi and Bluetooth-based systems are relatively low-cost but often suffer from signal instability and reduced accuracy in dense hospital structures (Zou et al., 2021; Möller et al., 2014). UWB offers higher precision but is cost-prohibitive and requires dedicated infrastructure (Xu et al., 2024). LiDAR, while effective for detailed 3D mapping, demands expensive hardware and significant post-processing (Zou et al., 2021).

By contrast, AR-based systems powered by vision-based SLAM enable real-time spatial awareness and require minimal physical setup. These systems support markerless tracking and dynamic visual overlays, making them well-suited for hospital layouts without major infrastructural modifications (Liu & Meng, 2020). The increasing availability of AR-capable smartphones and tablets makes this solution both portable and scalable.

**4.2 Cognitive Benefits of AR Navigation**

AR navigation reduces users' cognitive effort by embedding directional cues directly into their visual field. This minimizes the need for abstract reasoning, symbol decoding, or mental rotation—tasks that can be difficult under stress, especially in healthcare environments (Dong et al., 2021; Liu & Meng, 2020).

Seager and Fraser (2007) found that AR-based guidance improves spatial understanding and helps users process orientation cues more effectively. In emergency contexts, Ahn et al. (2024) observed that 88.6% of users preferred AR over traditional 2D maps due to its clarity and intuitiveness. These findings suggest that AR is particularly beneficial for individuals with limited spatial familiarity, cognitive impairments, or elevated anxiety.

**4.3 Navigation Efficiency and Real-Time Adaptability**

In addition to reducing mental workload, AR systems improve navigation accuracy and speed. Ahn et al. (2024) report that AR-guided users reoriented 25 seconds faster during unexpected detours and located target destinations 22 seconds quicker on average in emergency drills. These time gains are highly relevant in hospitals, where rapid and reliable navigation can directly impact care delivery.

Because AR interfaces can dynamically update paths based on user location and environmental changes, they are more flexible than static signage or printed maps. This adaptability ensures that users receive the most efficient, context-aware routes in real time.

**4.4 Accessibility, Scalability, and Cost Considerations**

One of AR's key advantages is its accessibility. Rather than relying on specialized infrastructure, AR navigation can operate on common mobile devices using vision-based SLAM. This makes deployment more feasible across different facilities, especially those lacking large IT budgets (Ahn et al., 2024).

AR has also been explored in assistive technologies for individuals with cognitive impairments, reinforcing its inclusive potential (Chang, Tsai, & Wang, 2008). Beyond user benefits, AR can reduce operational costs by decreasing staff time spent on providing directions. Zimring (1990) estimated that navigation-related inefficiencies could cost hospitals over $220,000 annually—an expense that AR has the potential to mitigate.

### 4.5 Lessons from Cultural and Museum Navigation

Museums, like hospitals, are spatially dense public environments serving a diverse range of users. Recent advances in AR and XR have been successfully applied to guide museum visitors through interactive, story-driven paths, improving spatial comprehension and engagement (Doukianou et al., 2020; Silva & Teixeira, 2020).

Chías et al. (2022) found that AR navigation systems significantly improved accessibility and orientation for users with disabilities in cultural spaces. These findings highlight transferable design principles: hospitals can similarly adopt AR to enhance wayfinding, particularly for vulnerable groups, by delivering personalized, multimodal, and context-sensitive guidance.

## 5 RESEARCH QUESTIONS AND IDENTIFIED GAPS

Despite increasing interest in improving hospital wayfinding, most existing solutions remain grounded in static signage systems or basic 2D digital maps. These approaches, while functional, often fail to address the cognitive, emotional, and spatial challenges faced by users in complex healthcare environments. Meanwhile, immersive technologies such as Augmented Reality (AR) have shown potential in domains like museum navigation and cultural heritage sites, yet their application in hospitals remains limited and underexplored.

This study addresses this gap by examining the cognitive, behavioral, and emotional impacts of AR-based navigation in comparison to traditional methods in a real-world hospital setting.

### 5.1 Research Questions

This study is guided by the following research questions:

1. **Cognitive Load:** How does AR-based navigation compare with traditional paper-based methods in terms of cognitive workload during hospital wayfinding tasks?

2. **Navigation Efficiency:** Does AR guidance improve navigation efficiency—measured by task completion time, path accuracy, and error reduction?

**Emotional Experience:** Can AR-based navigation reduce user anxiety and enhance the overall navigation experience for patients and visitors?

## 5.2 Identified Gaps in Prior Research

- **Limited focus on AR in healthcare wayfinding:** While prior studies have examined AR for education, tourism, or exhibition settings, few have explored its role in healthcare environments—particularly under stress-inducing conditions common in hospitals.
- **Lack of cognitive and emotional evaluation:** Most hospital wayfinding research emphasizes visual signage or digital map design, with minimal attention to how immersive guidance impacts cognitive load, spatial understanding, or emotional comfort.
- **Underexplored user diversity:** There is limited research examining how navigation systems serve users with varying cognitive abilities, familiarity with hospital settings, or levels of digital literacy.

This study contributes to closing these gaps by conducting an empirical comparison of AR and paper-based navigation in a simulated real-world hospital context, focusing on measurable performance, perceived workload, and user experience outcomes.

## 6. METHOD

To provide a structured overview of our research process, Figure 01 presents the study framework, outlining the problem motivation, system conditions, experimental procedure, data collection methods, and evaluation approach used in this study.

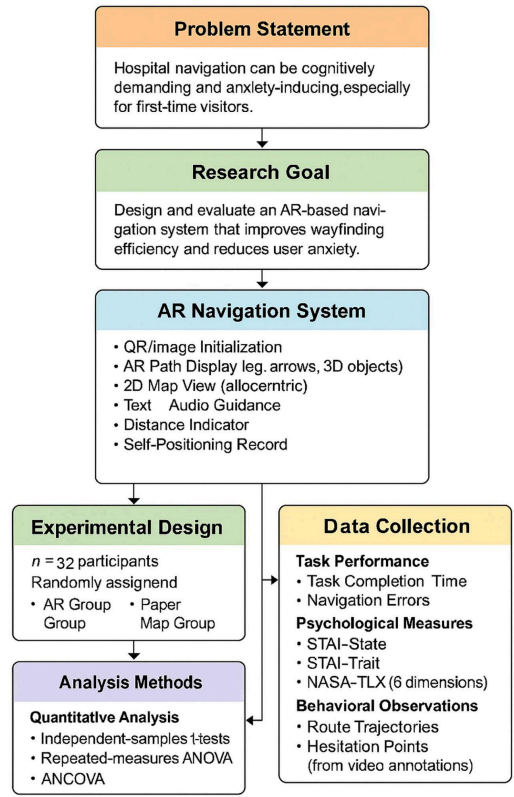

*Figure 01*: *Research framework for the AR navigation study*

## 6.1 Participants

Thirty-two adult participants (19 female, 13 male; ages 19–30, $M = 28.4$) were recruited from the university community to participate in an indoor navigation experiment at the University of Michigan Hospital. All reported daily smartphone use and no prior experience with augmented reality (AR) navigation systems. Participants were randomly assigned to either an AR navigation condition ($n = 16$) or a paper map condition ($n = 16$), with randomization stratified by gender. All participants provided informed consent under IRB Protocol HUM00271415（see Appendix A for full survey items）and completed a demographic survey and baseline anxiety assessment.

## 6.2 Study Design and Procedure

**Experimental Design**

A between-subjects design was used to compare AR-based navigation with traditional 2D paper map navigation. The task involved traveling from the 2nd-floor parking entrance of the Taubman Center to the Pathology Center and returning via an alternate route to the starting point. This return path differed from the initial route to encourage broader spatial exposure and test route integration. Participants traveled a 280-meter route from the parking entrance to the destination and then returned via an alternate path, totaling roughly 560 meters of indoor navigation on the same floor.s.

**Study Setting**

The University of Michigan Hospital was chosen for its multi-level structure and clinic-dense layout. It includes standard wayfinding infrastructure such as printed maps and directional signage, enabling realistic comparison with the AR system. The research team's familiarity with the facility allowed for standardized route planning.

**Procedure**

Each session lasted approximately 15–20 minutes and followed a standardized protocol:

- **Consent and Baseline Assessment:**
  Participants completed informed consent, demographic surveys, and baseline STAI-State and STAI-Trait questionnaires. (see Appendix B, C and D  for full survey items)
- **Condition Familiarization:**

AR group: Received brief (~1 minute) instruction on interpreting AR overlays.

Paper map group: Received a short briefing on the 2D map layout and symbols.

All participants were instructed not to seek external assistance during navigation.

**Outbound Navigation Task:**

Participants were asked to imagine they were heading to a medical appointment and to complete the task efficiently, without time pressure. They navigated independently from the 2nd-floor parking entrance from Taubman Center to the pathology Center.

**Post-Task 1 Assessment:**

Participants completed NASA-TLX(see Appendix E  for full survey items) and STAI-State questionnaires.

**Return Navigation Task:**

After a short break (~2 minutes), participants navigated back to the starting point using the same method.

- **Post-Task 2 Assessment:**

  Participants again completed NASA-TLX and STAI-State assessments.

- **Qualitative Interview and Route Sketching:**

  Each participant participated in a ~10-minute semi-structured interview, reflecting on their navigation experience, interface usability, and spatial confidence. They also sketched the navigated route from memory to assess spatial understanding.

## 6.3 System Conditions

**AR Navigation Interface**

Participants in the AR condition interacted with a mobile-based augmented reality interface (Figure 02) that provided real-time spatial guidance. The interface included directional arrows and turn indicators displayed through the device's camera view, along with an optional top-down minimap to support orientation.

In the control condition, participants were provided with a printed 2D floor map (Figure 03), identical to those distributed at hospital information desks and used by real patients. The map included labeled rooms, hallways, and orientation icons (e.g., north arrows, elevators). Participants were instructed to rely solely on the map to complete the tasks without external assistance or intervention..

In both conditions, participants received brief (~1 minute) instruction before the task and completed the route twice—first from the Taubman center's parking entrance to the pathology department, and then returning via an alternate path. Each session was recorded using a chest-mounted GoPro Hero4.

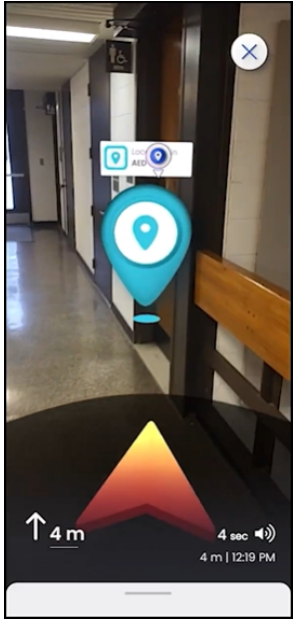
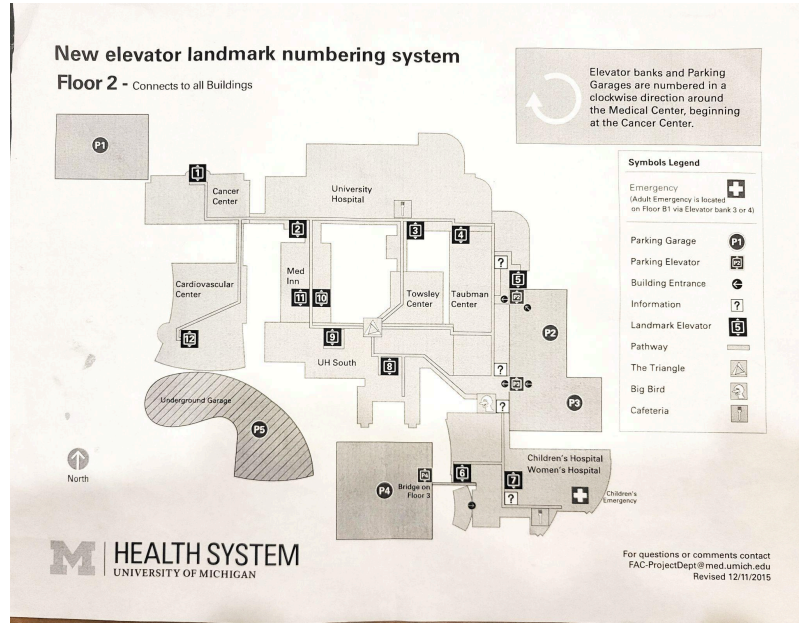

***Figure 02.*** *AR navigation interface*          ***Figure 03.*** A printed 2D hospital map

**Additional AR Task (Paper Map Group Only)**

After completing both tasks using the paper map, participants in the control group (paper group) completed an additional route (~200 meters) using the AR interface. This route was of similar complexity but previously unexplored. After a brief (~2-minute) AR tutorial, participants navigated the route and completed a comparative satisfaction survey covering ease of use, anxiety, intuitiveness, and system preference.

**6.4 Measures**

**Objective Performance Metrics**

- **Task Completion Time:** Time (in seconds) to reach each destination.
- **Path Accuracy:** We calculated path accuracy as the percentage deviation from the predefined optimal route. This metric penalizes extra walking due to wrong turns, hesitations, or inefficient routing and provides a normalized measure of navigational efficiency. The formula used is:

$$\text{Path Accuracy (\%)} = \left( \frac{\text{Optimal Path Length}}{\text{Actual Path Length}} \right) \times 100$$

Manual annotations of turning points, route deviations, and hesitation durations were derived from chest-mounted GoPro recordings. Annotations were temporally aligned with the sketched trajectory data to qualitatively examine behavioral differences across conditions.

- **Navigation Errors:** Number of wrong turns or hesitation events (>5 seconds).

**Subjective Measures**

- **Cognitive Workload:** Measured via the NASA Task Load Index (NASA-TLX; Hart & Staveland, 1988), covering mental, physical, and temporal demand, effort, performance, and frustration.
- **Anxiety:** Measured using the STAI-State scale (Spielberger, 1983) before and after each navigation task.

**Qualitative Measures**

- **Interviews:** Post-navigation semi-structured interviews explored user experiences, confidence, and interface impressions.
- **Sketch-Based Spatial Recall:** Participants sketched the navigated route from memory to evaluate spatial understanding.(see appendix F )

## 7. AR WAYFINDING SYSTEM DESIGN PRINCIPLES AND TECHNICAL IMPLEMENTATION

### 7.1 Theoretical Foundations of Wayfinding

Navigating large-scale, unfamiliar environments—such as hospitals—places significant cognitive demands on users, especially under stress or time constraints. Prior research in spatial cognition and environmental psychology identifies five essential components of effective wayfinding: (1) identifying one's current location, (2) recognizing the intended destination, (3) selecting an appropriate route, (4) confirming arrival, and (5) returning to the point of origin (Ulrich et al., 2010; Carpman & Grant, 2016).

Breakdowns in any of these steps can result in confusion, increased anxiety, or delays in reaching medical appointments—consequences that are particularly problematic in healthcare settings. These five wayfinding tasks thus offer a useful theoretical framework for evaluating and guiding the design of spatial navigation systems in hospitals.

Building upon this foundation, we developed an AR-based navigation system aimed at directly supporting each of these cognitive processes through intuitive, real-time, and context-aware design strategies.

## 7.2 AR Design Principles for Hospital Navigation

The system was informed by interdisciplinary insights from human-computer interaction (HCI), spatial cognition, and healthcare design. Three core design principles guided the interface and user experience:

### 7.2.1 Embedded Visual Guidance and Real-Time Anchoring

While our system does not employ a fully immersive egocentric perspective (e.g., via head-mounted AR displays), it presents a camera-aligned, forward-facing view that mimics key aspects of egocentric guidance. Prior research has demonstrated that such alignment—whether through handheld or immersive AR—can significantly reduce the cognitive load associated with spatial translation (Seager & Fraser, 2007; Xu et al., 2023).

By embedding directional cues directly within the user's real-time view of the environment, the system supports intuitive decision-making and continuous orientation. This reduces the need to interpret abstract map symbols or recall previous instructions, which is especially beneficial in complex and emotionally charged environments like hospitals. Real-time visual anchoring helps mitigate spatial anxiety and promotes smoother, more confident navigation (Nori et al., 2023).

### 7.2.2 Minimalist Interface for Cognitive Load Reduction

The interface design emphasizes simplicity and clarity. To prevent cognitive overload, only essential visual elements are displayed, including a directional arrow, destination label, and minimal contextual markers. Unnecessary graphics or animations were excluded to maintain user focus. This minimalist design aligns with established cognitive load reduction principles and is especially suitable for environments where users may be under stress or time pressure (Nori et al., 2023).

### 7.2.3 Optional 2D Overview for Situational Awareness

In addition to the real-time AR interface, the system includes a toggleable full-scale 2D overview map. This allocentric map provides users with a global understanding of the floor layout and their real-time location within it. This feature directly supports the first stage of wayfinding—identifying one's current position—as defined in spatial cognition literature(Seager & Fraser, 2007; Xu et al., 2023). By bridging egocentric and allocentric perspectives, the 2D map allows users to develop situational awareness and select navigation strategies based on personal cognitive preferences.

### 7.3 Technical Implementation

The AR navigation system was developed using Unity and deployed on Android smartphones. Visual spatial anchoring was enabled via ARwayKit, which provides marker-less SLAM (Simultaneous Localization and Mapping) capabilities. This allowed the system to align virtual content—such as arrows, labels, and turn indicators—with real-world surfaces without requiring external markers or beacons.

To ensure accurate environmental mapping, the hospital corridor was first captured using LiDAR-based 3D scanning（Figure 4）. These scans were processed into a spatial mesh, which served as the basis for placing coordinate-anchored digital overlays within the Unity scene. The configuration ensured that all AR content remained spatially stable across devices and sessions.

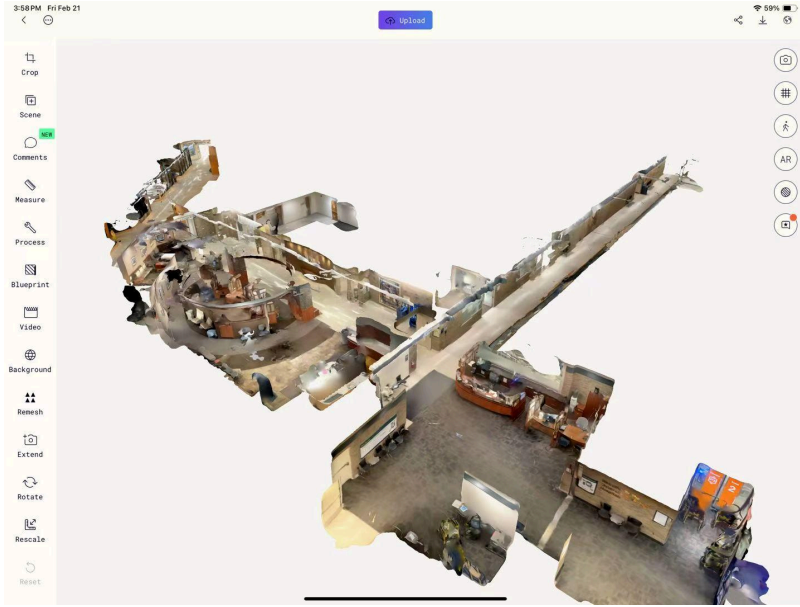

***Figure 4.*** *LiDAR-based 3D scanning for Taubman Center in University of Michigan Hospital*

The modular AR interface supported several navigation-related features, including（Figure 5）:

- **QR Code or Image-Based Initialization:** Users activate the navigation system by scanning a location-specific QR code or reference image to establish their starting point.
- **Route Selection:** Users select a predefined route or choose from multiple destinations.
- **AR Path Display:** Visual cues (e.g., arrows or 3D objects) appear directly in the environment, anchored to real-world surfaces.
- **Real-Time 2D Map Integration:** A toggleable allocentric map shows the user's current position and orientation at any time.
- **Multimodal Support:** Text and audio instructions are offered to enhance accessibility for users with different cognitive preferences.
- **Distance Indicator:** A floating label indicates the remaining distance to the destination (in meters), which updates dynamically.
- **Self-Positioning Record:** The system logs user paths for analysis and can assist users in returning to the origin.
- **Low-Clutter Interface:** The interface is optimized for quick comprehension and minimal stress under hospital conditions.

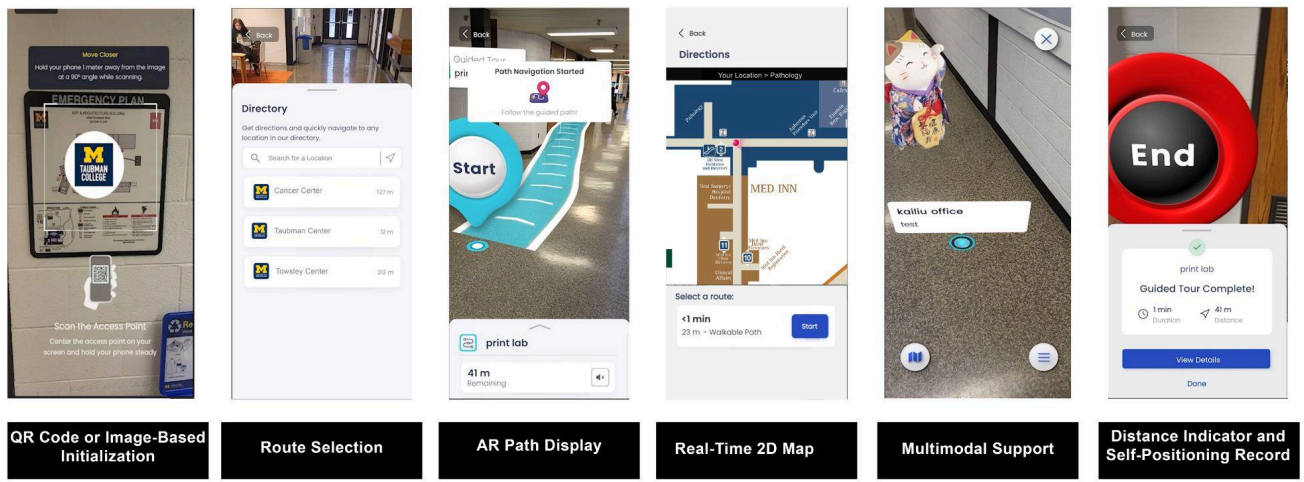

| QR Code or Image-Based Initialization | Route Selection | AR Path Display | Real-Time 2D Map | Multimodal Support | Distance Indicator and Self-Positioning Record |

***Figure 5.*** *Technical architecture showing Unity–ARwayKit pipeline for hospital AR deployment.*

The system was field-tested at the University of Michigan Hospital. The chosen routes incorporated changes in floor level, intersections, and common wayfinding obstacles such as signage confusion and hallway convergence zones.

## 7.4 Integration with Experimental Evaluation (Figure 06)

The design of the AR system directly informed the experiment protocol described in Chapter 6. Based on the system's interface and features, the following hypotheses were formulated and empirically tested:

- **H1:** AR navigation will enable faster task completion than traditional paper maps.
- **H2:** AR users will commit fewer navigation errors or wrong turns.
- **H3:** Participants using AR will report lower perceived workload (NASA-TLX).
- **H4:** AR guidance will reduce spatial anxiety as measured by STAI-State.

- **H5:** Participants using AR will report higher satisfaction and preference for future use.

| AR Feature | H1<br>(Efficiency) | H2<br>(Accuracy) | H3<br>(Cognitive Load) | H4<br>(Anxiety) | H5<br>(Satisfaction & Preference) |
|---|---|---|---|---|---|
| Real-time directional arrows | ✅ | ✅ | | | ✅ |
| Minimalist, low-clutter interface | | | ✅ | ✅ | ✅ |
| 2D overview map (allocentric) | ✅ | | ✅ | | ✅ |
| Voice and text instructions | | | ✅ | ✅ | ✅ |
| Distance-to-destination indicator | ✅ | | | | ✅ |
| Re-routing and real-time updates | ✅ | ✅ | | | ✅ |

***Figure 06 .*** *Mapping between AR features and experimental hypotheses.*

These hypotheses were evaluated using a mixed-methods approach, including behavioral tracking, quantitative assessments, and post-task interviews to assess usability, performance, and user experience.

## 8. RESULTS AND ANALYSIS

### 8.1 Quantitative Findings: Navigation Performance and User Experience

**Task Completion Time (Figure 7)**

Participants in the AR condition completed navigation tasks significantly faster (M = 574.19 s, SD = 4.30 s) than those using the paper map (M = 673.74 s, SD = 7.17 s). An independent-samples t-test confirmed that this difference was statistically significant, t(30) = -7.28, p < .001, indicating that AR guidance improved overall navigation efficiency.

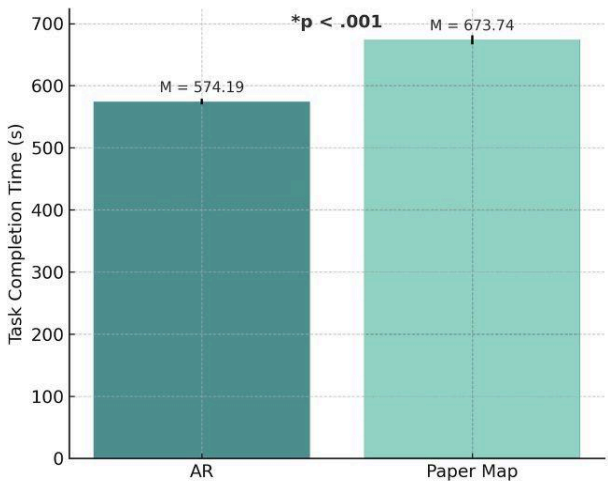

Figure 1. Task Completion Time by Navigation Method

*Figure 07 :*Task Completion Time

**Navigation Errors (Figure 8)**

Participants in the AR group made significantly fewer navigation errors (M = 0.75, SD = 0.88) than those in the paper map group (M = 2.44, SD = 0.94), t(30) = -5.71, p < .001. This suggests that AR support led to more accurate decision-making during navigation.

**Path Accuracy (Figure 9)**

Adherence to the optimal route was significantly higher in the AR group (M = 93.21%, SD = 1.50%) compared to the paper map group (M = 88.74%, SD = 1.85%), t(30) = 4.38, p < .001. Heatmap visualizations further support this finding, showing that AR users followed more streamlined and direct paths.

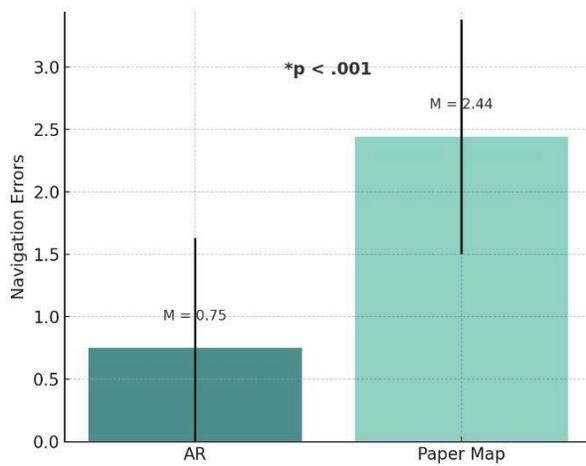

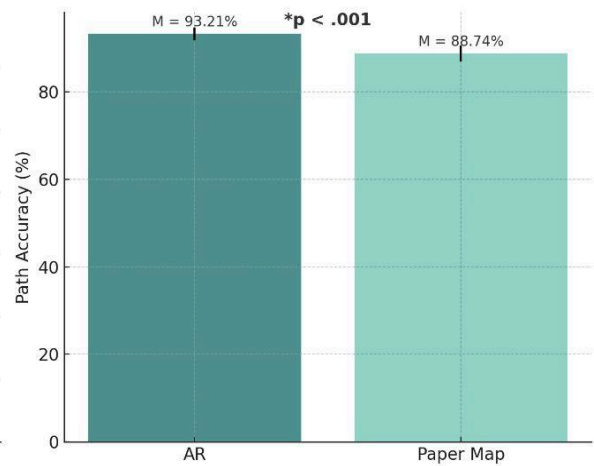

***Figure 8****: Navigation Errors*  ***Figure 9****: Path Accuracy*

## 8.2 Cognitive Load and Anxiety Assessment

**NASA-TLX Workload Scores (Figure 10)**

We measured subjective workload using the NASA-TLX scale across six dimensions. A series of independent samples t-tests revealed significant differences between the AR and Paper groups in several dimensions. AR participants reported lower mental demand (M = 15.9, SD = 11.7) than those using paper maps (M = 33.4, SD = 11.0), t(30) = -4.49, p < .001. Significant differences were also observed in effort (M = 36.2, SD = 8.9 vs. 49.8, SD = 11.6;t(30) = -3.80, p < .001) and performance ratings (M = 90.4, SD = 4.9 vs. 73.4, SD = 10.1; t(30) = 6.24, p < .001), with AR users reporting higher subjective performance.

No significant differences were found in physical demand (t(30) = 0.05, p = .96) or temporal demand (t(30) = -0.34, p = .74), suggesting that both groups experienced comparable levels of physical and time-related load. These results indicate that AR-based navigation reduces cognitive and emotional workload, but does not substantially affect perceived physical or temporal demand.

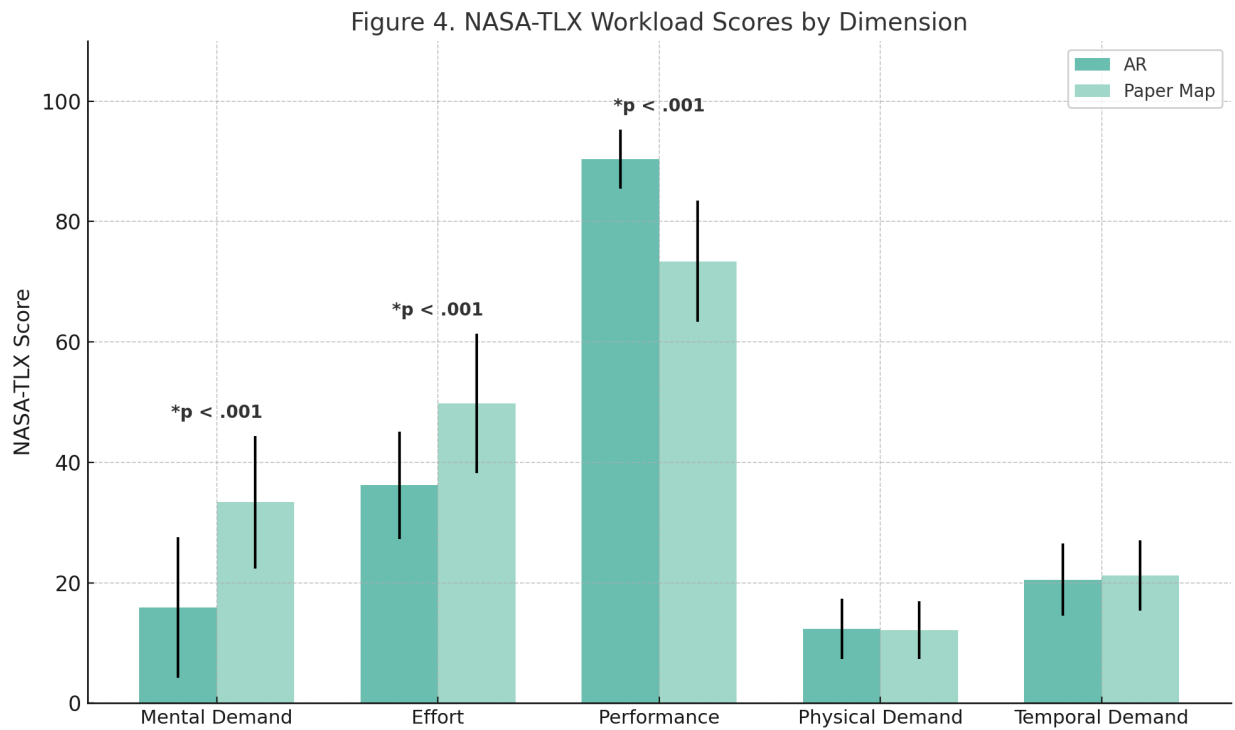

Figure 4. NASA-TLX Workload Scores by Dimension

*Figure 10 : NASA-TLX Workload Scores*

## State Anxiety (STAI-State, Figure 11)

State anxiety was measured at three time points: before navigation (baseline), after reaching the destination (post-task 1), and after returning to the starting point (post-task 2). Both AR and Paper groups showed comparable baseline anxiety (AR: M = 44.9, SD = 3.3; Paper: M = 43.8, SD = 3.9), and no significant differences were observed in their STAI-Trait scores (AR: M = 39.9, SD = 5.0; Paper: M = 39.8, SD = 4.7), indicating similar general anxiety tendencies across groups. Independent samples t-tests confirmed no significant difference in STAI-Trait (t(30) = 0.07, p = .94) or baseline STAI-State (t(30) = 0.93, p = .36).

A repeated-measures ANOVA revealed a significant condition × time interaction, F(2, 60) = 5.93, p = .004. Post-hoc comparisons showed that participants using AR experienced a notable decrease in state anxiety after both navigation phases (Mid: M = 41.2, SD = 3.4; Post: M = 40.5, SD = 3.1) compared to the Paper group (Mid: M = 44.9, SD = 3.9; Post: M = 45.1, SD = 3.8). These differences remained significant even when controlling for Trait anxiety using ANCOVA (F(2, 89) = 5.93, p = .004), confirming that AR-based navigation effectively reduced anxiety beyond baseline individual differences.

These findings suggest that real-time AR guidance not only improved task performance but also alleviated emotional stress during complex hospital wayfinding—particularly benefiting users with higher trait anxiety.

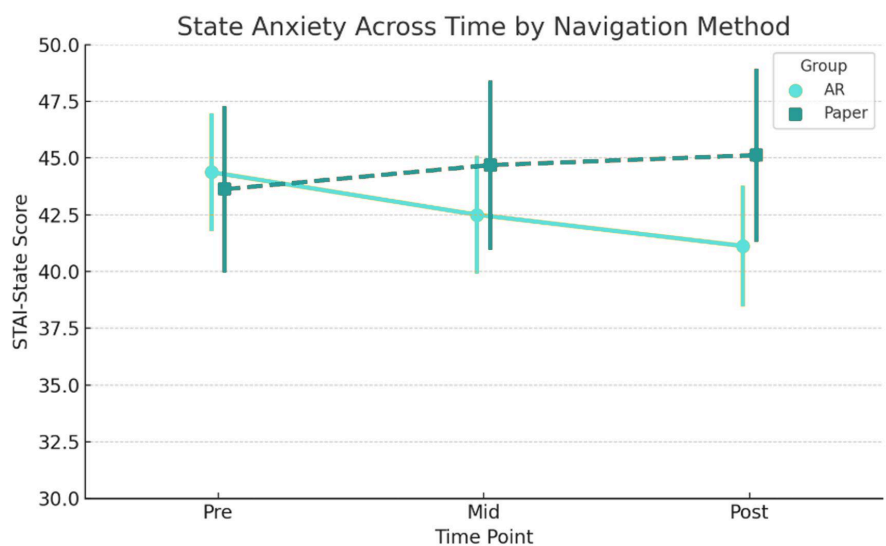

***Figure 11 :*** *State Anxiety Across Time by Navigation Method*

**Figure11.** Mean state anxiety scores across three time points (Pre, Mid, and Post) for participants using AR-based and paper-based navigation methods. Error bars represent standard deviations. AR users showed a consistent decrease in anxiety over time, while paper map users exhibited a slight increase.

## 8.3 Spatial Behavior: User Routes and Hesitation Mapping

### Participant Trajectory Mapping (Figure 12)

To better understand navigational behavior, we visualized the spatial trajectories of participants in both the AR and Paper Map conditions. Overlaid on the hospital floorplan, these paths reveal clear differences in route directness and hesitation patterns. Participants in the AR group generally followed more consistent and streamlined routes, while Paper Map users often exhibited detours, backtracking, or pauses—marked as hesitation points on the map. These patterns suggest that AR guidance contributed to more confident and efficient wayfinding behaviors.

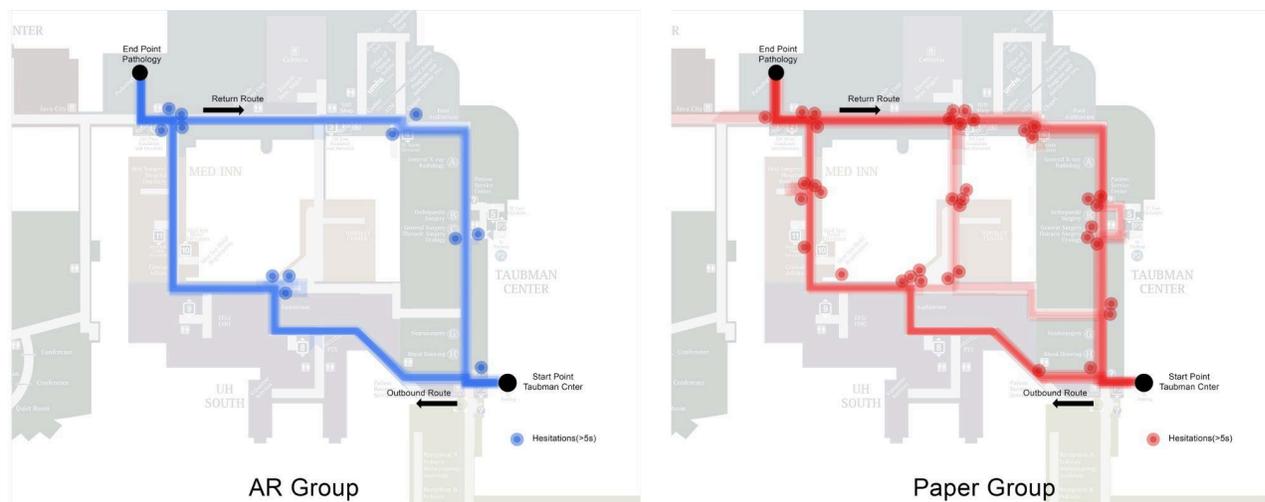

***Figure 12*** *: Participant Trajectory Map with Hesitation Points*

**Route Comparison by Condition**

While we did not conduct quantitative space syntax analysis, we qualitatively compared the distribution and structure of selected routes between groups. AR-guided participants tended to follow spatially intuitive corridors and highly visible zones, aligning with what space syntax literature would describe as "integrated" paths. In contrast, Paper Map users more frequently entered less structured areas or dead ends, which may have contributed to increased navigation errors and uncertainty.

**8.4 User Satisfaction and Preference**

**Initial Satisfaction Ratings**

We assessed user satisfaction across navigation conditions using a 5-point Likert scale. Participants who used the AR system reported significantly higher initial satisfaction (M = 4.52, SD = 0.47) than those who used paper maps (M = 2.72, SD = 0.75). A Mann-Whitney U test confirmed this difference was significant, U = 248.0, p < .001

**Post-AR Experience Rating**

To evaluate satisfaction changes, paper map users were asked to re-rate their experience after using the AR system. Satisfaction increased significantly following AR exposure (M = 4.56, SD = 0.35), Z = 0.0, p < .001 (Figure 9), indicating a strong positive impact of AR even among users initially unfamiliar with the technology. Participants commonly noted the system's ease of use and reduction in mental effort during qualitative feedback.

**Willingness for Future Use**

When asked about future preferences, 94% of participants indicated they would choose AR-based navigation in future hospital visits, while only 31% expressed willingness to reuse paper maps. These results suggest that AR interfaces not only enhance immediate satisfaction but also increase long-term preference and adoption potential.

### 8.4.1 Sketch-Based Spatial Memory Assessment（Figure 13）

To further evaluate participants' internal spatial understanding, all users were asked to draw the routes they had navigated from memory after completing the tasks. This qualitative sketching activity served as an additional lens for assessing cognitive mapping.

Participants in the paper map condition generally produced more detailed and accurate sketches, including recognizable landmarks, intersection points, and overall route structure. Their drawings reflected a stronger grasp of the building's layout, suggesting that reliance on 2D maps may promote the formation of configurational spatial knowledge.

In contrast, AR navigation users tended to sketch simplified, sequential paths, often focusing on the order of turns rather than the spatial relationships between destinations. Their route representations were more procedural, indicating less emphasis on constructing a holistic mental model of the environment.

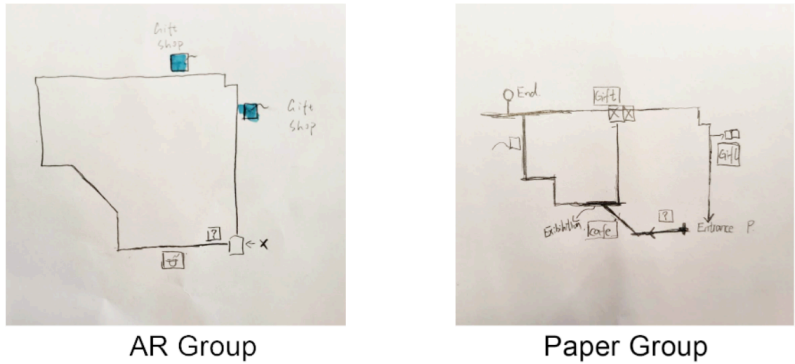

| AR Group | Paper Group |

***Figure 13*** *: Selected  Spatial Memory sketch*

These findings support prior research suggesting that traditional map-based navigation engages users in active spatial inference and mental mapping, while AR systems, by offering continuous visual guidance, may reduce the need for such cognitive engagement (Passini et al., 2000; Gardony et al., 2018). While this enhances short-term efficiency and reduces cognitive load, it may limit long-term spatial memory development.

This trade-off highlights a key design consideration: AR navigation systems may benefit from incorporating contextual or landmark-based cues to better support incidental learning and spatial orientation over time.

## 8.5 Qualitative Insights: Perception and Spatial Learning

Semi-structured interviews revealed key themes that align with quantitative findings:

**Confidence and Control:**
 "With AR, I always felt I knew exactly where to go."

**Ease of Use:**
 "The AR arrows were straightforward and very intuitive."

**Contrast with Paper Maps:**
 "The paper map was frustrating… but AR completely changed my experience."

AI-Enhanced AR Experiences

Beyond navigation, participants envisioned a more interactive, intelligent system—one that not only guides but also communicates. Several participants described a desire for an AI-driven AR assistant that could offer real-time guidance, check-in support, and emotional reassurance during stressful hospital visits. One participant noted:

> "If there was an avatar who greeted me and told me where to wait or when to go, I wouldn't feel so lost."

This suggests strong user interest in context-aware, emotionally responsive systems. Integrating AI agents within AR interfaces could enable real-time interaction, predictive guidance based on appointments, and even anxiety-reducing dialogue—all of which contribute to a more seamless and human-centered hospital journey.

These comments reflect a consistent preference for AR and reinforce its value in supporting orientation, emotional regulation, and spatial comprehension.

### 8.6 Summary of Key Findings

The multi-method results consistently demonstrate the effectiveness of AR-based wayfinding in hospital environments. Quantitative metrics highlight improvements in efficiency, error reduction, and reduced cognitive and emotional load. Spatial syntax analyses further validate AR's alignment with intuitive environmental affordances. Qualitative feedback confirms user satisfaction and experiential benefits.

Taken together, these findings underscore AR navigation's potential to enhance hospital navigation by addressing both functional and emotional user needs in complex spatial settings.

## 9. DISCUSSION

### 9.1 Key Findings

This study investigated how handheld AR navigation compares with traditional paper maps in a real-world hospital environment. Our results indicate that participants using the AR system significantly outperformed those using paper maps across multiple dimensions.

Quantitatively, AR users completed navigation tasks more quickly, committed fewer errors, and adhered more closely to optimal routes. They also reported lower perceived cognitive workload (NASA-TLX) and reduced situational anxiety (STAI-State), highlighting the benefits of real-time, camera-aligned directional cues in reducing the cognitive burden of navigating unfamiliar hospital environments.

Participants also expressed significantly higher satisfaction with the AR system. Post-task questionnaires revealed that 94%of participants indicated a strong willingness to use AR navigation in future hospital visits, compared to only 31%for paper maps. Paper map users who later experienced AR navigation described it as more intuitive, less mentally taxing, and generally preferable. This reinforces the potential for AR-based systems to be adopted in real-world healthcare settings.

However, a notable trade-off emerged in spatial recall and environmental awareness. Participants in the paper map condition performed better on route sketching and demonstrated a stronger understanding of the hospital's spatial layout during interviews. This suggests that while AR provides navigational ease, it may inhibit the development of long-term spatial knowledge—consistent with prior findings on the cognitive offloading effects of turn-by-turn systems (Ishikawa et al., 2008).

In summary, AR navigation systems provide clear advantages in task performance, user satisfaction, and stress reduction, though designers must remain attentive to their potential impact on spatial learning.

**9.2 Interpretation of Results**

The performance benefits observed in the AR condition support existing literature that highlights how egocentric-style guidance—whether through head-mounted displays or handheld devices—can reduce the need for abstract spatial transformations, leading to more efficient navigation (Xu et al., 2023; Seager & Fraser, 2007).

Unlike paper maps, which require users to constantly translate a top-down layout into first-person movement decisions, the AR interface overlays directional cues directly onto the user's live camera view. This approach reduces memory load and mental mapping effort (Nori et al., 2023), as evidenced by significantly lower NASA-TLX scores. Participants in the AR condition also reported less situational anxiety, likely due to the presence of clear, forward-facing guidance cues that reduced ambiguity and uncertainty during navigation.

However, the enhanced ease of AR may come at the cost of reduced spatial learning. Participants who used paper maps engaged more actively with the overall layout and structure of the hospital, which likely contributed to their superior performance in spatial memory tasks. This finding echoes the classic trade-off of cognitive offloading: by reducing the need for internal processing, guidance systems can improve efficiency but potentially undermine users' situational awareness and environmental learning (Ishikawa et al., 2008).

### 9.3 Design Implications

### Reducing Cognitive Load and Anxiety

Handheld AR navigation systems with camera-aligned cues can significantly reduce cognitive burden and emotional stress during navigation. In hospital settings—where patients and visitors may be anxious, time-constrained, or unfamiliar with their surroundings—minimizing the need for abstract map interpretation is especially beneficial. Design strategies should emphasize clear, intuitive overlays such as forward-facing arrows, floating markers, and progress indicators. These cues should be visually salient yet unobtrusive to support cognitive clarity under pressure.

### Mitigating the Spatial Learning Trade-Off

To address the potential decline in spatial knowledge acquisition, designers should augment AR systems with environmental anchoring strategies. For example, incorporating landmark-based prompts (e.g., "Turn right after the pharmacy") can encourage users to engage with their surroundings, supporting incidental learning. Allowing momentary switches to a top-down view or showing faded traces of previous paths may also aid spatial orientation. Such hybrid designs can help users navigate efficiently while gradually building mental maps of unfamiliar spaces.

### Inclusive Design for Neurodiverse Users

Our findings underscore the value of adaptive AR systems that cater to users with diverse cognitive styles and abilities. Handheld AR systems reduce the demand for spatial transformation and abstract reasoning—tasks that can be challenging for users with ADHD, anxiety, or mild cognitive impairments. Interface customization options—such as audio instructions, simplified visual elements, or adjustable pacing—can improve accessibility for neurodiverse populations while benefiting all users through enhanced clarity and reduced overload.

## 9.4 Limitations

This study has several limitations that inform opportunities for future work. First, the experiment was conducted at a single hospital site (University of Michigan Hospital), using one predefined route, which may limit the generalizability of the results to other architectural layouts or signage systems. The participant sample was relatively small (N = 32) and drawn primarily from a young, tech-savvy university population, potentially biasing results in favor of digital systems like AR. While our findings showed short-term benefits in reducing anxiety and improving navigation performance, participants interacted with the system in only a single session, and novelty effects may have influenced their perceptions. We also measured spatial memory immediately after the task, without assessing long-term retention. Furthermore, our spatial awareness assessment—based on route sketching—captures only one dimension of cognitive mapping; future work should explore delayed recall or free re-navigation. The use of STAI-State to assess anxiety provides only a single post-task snapshot; real-time or physiological measures could better capture where anxiety peaks occur, such as at junctions or decision points. Finally, the AR system's performance was constrained by current hardware and visual SLAM limitations, occasionally causing tracking drift or misaligned overlays—issues that may be reduced as AR technologies continue to advance.

## 9.5 Future Work

We envision several future directions that can extend this research and inform the next generation of inclusive navigation technologies. First, longitudinal studies are essential to assess whether repeated use of AR navigation enhances users' spatial knowledge or reinforces reliance on passive guidance. An open question remains: does AR improve navigational performance while undermining configurational spatial memory over time? We also suggest exploring adaptive interfaces that tailor support based on user traits or in-the-moment behaviors—such as switching between egocentric and allocentric views based on navigation performance or anxiety indicators—thus balancing cognitive ease with environmental learning.

Building tighter integration with hospital infrastructure represents another promising avenue. Coupling AR systems with real-time appointment updates, dynamic signage, or IoT networks could enhance navigational precision while streamlining the overall care experience. In addition, we advocate testing AR tools in high-pressure or time-sensitive situations (e.g., emergency care, evacuation drills) to assess how stress levels interact with AR usability, which remains underexplored in healthcare HCI.

Critically, AR navigation could provide disproportionate benefits in resource-constrained hospitals, where traditional signage is limited and staff are overextended. As highlighted by Upadhyay et al. (2022), patients in overcrowded hospitals often struggle with unclear pathways, exacerbating anxiety and appointment delays. Future research should evaluate AR deployment in such contexts—particularly in developing regions—to assess its potential in reducing care inequities.

Finally, broadening AR application domains (e.g., airports, public transit hubs, university campuses) and comparing handheld versus head-worn displays will help define modality-specific advantages. Collectively, these directions reinforce our broader goal: to develop AR navigation systems that are not only functionally efficient but also cognitively supportive, emotionally inclusive, and scalable across diverse environments.

## 10  CONCLUSION

This study evaluated an AR-based indoor navigation system in a complex hospital setting, comparing it against traditional paper maps. Results showed that AR significantly reduced task completion time, navigation errors, cognitive workload, and situational anxiety. Participants also expressed a strong preference for AR in future use, indicating its practical potential in real-world healthcare environments.

Our system was designed based on foundational wayfinding theory and spatial cognition research. By combining embedded directional cues, a minimalist interface, and a toggleable allocentric 2D map, the AR system supported key user actions such as locating oneself, selecting routes, and confirming arrival. This design also benefited users with diverse cognitive styles, including those prone to stress or difficulty interpreting abstract maps.

While AR improved short-term navigation efficiency, findings revealed a trade-off in spatial learning. Paper map users demonstrated stronger spatial memory, suggesting that AR's cognitive offloading may limit long-term environmental awareness. To mitigate this, we recommend incorporating landmark-based cues and flexible perspective-switching to balance guidance with learning.

This work contributes actionable design principles for inclusive, context-aware AR systems in healthcare. As AR-capable devices become increasingly widespread, hospitals and designers have a valuable opportunity to enhance accessibility and reduce anxiety for diverse patient populations. Future work should explore adaptive AR interfaces, deeper integration with hospital infrastructure, and applications in high-stress or resource-limited environments.

# Appendix

Appendix C

## STAI-Trait (Trait Anxiety Inventory) – Bilingual Version

Instructions:
 Below is a list of statements. Read each statement and choose the response that best describes how you generally feel. Circle the appropriate number.

 说明：
以下是一些描述您通常感受的陈述。请阅读每一项，并圈出最符合您一般感受的选项。

1. I feel pleasant. / 我感到愉快。
 1 = Almost Never / 几乎从不 2 = Sometimes / 有时 3 = Often / 经常　4 = Almost Always / 几乎总是

2. I tire quickly. / 我很快就会感到疲倦。
 1 = Almost Never / 几乎从不 2 = Sometimes / 有时 3 = Often / 经常　4 = Almost Always / 几乎总是

3. I feel like a failure. / 我觉得自己是个失败者。
 1 = Almost Never / 几乎从不 2 = Sometimes / 有时 3 = Often / 经常　4 = Almost Always / 几乎总是

4. I feel rested. / 我感到休息得很好。
 1 = Almost Never / 几乎从不 2 = Sometimes / 有时 3 = Often / 经常　4 = Almost Always / 几乎总是

5. I am 'calm, cool, and collected.' / 我感到冷静、沉着、有条理。
 1 = Almost Never / 几乎从不 2 = Sometimes / 有时 3 = Often / 经常　4 = Almost Always / 几乎总是

6. I feel that difficulties are piling up so that I cannot overcome them. / 我觉得困难堆积如山，无法克服。

1 = Almost Never / 几乎从不 2 = Sometimes / 有时 3 = Often / 经常　4 = Almost Always / 几乎总是

7. I worry too much over something that really doesn't matter. / 我为一些其实无关紧要的事情担心太多。

1 = Almost Never / 几乎从不 2 = Sometimes / 有时 3 = Often / 经常　4 = Almost Always / 几乎总是

8. I am happy. / 我很开心。

1 = Almost Never / 几乎从不 2 = Sometimes / 有时 3 = Often / 经常　4 = Almost Always / 几乎总是

9. I have disturbing thoughts. / 我有一些困扰我的想法。

1 = Almost Never / 几乎从不 2 = Sometimes / 有时 3 = Often / 经常　4 = Almost Always / 几乎总是

10. I lack self-confidence. / 我缺乏自信。

1 = Almost Never / 几乎从不 2 = Sometimes / 有时 3 = Often / 经常　4 = Almost Always / 几乎总是

11. I feel secure. / 我感到很有安全感。

1 = Almost Never / 几乎从不 2 = Sometimes / 有时 3 = Often / 经常　4 = Almost Always / 几乎总是

12. I make decisions easily. / 我很容易做决定。

1 = Almost Never / 几乎从不 2 = Sometimes / 有时 3 = Often / 经常　4 = Almost Always / 几乎总是

13. I feel inadequate. / 我觉得自己不够好。

1 = Almost Never / 几乎从不 2 = Sometimes / 有时 3 = Often / 经常　4 = Almost Always / 几乎总是

14. I am content. / 我感到满足。

1 = Almost Never / 几乎从不 2 = Sometimes / 有时 3 = Often / 经常　4 = Almost Always / 几乎总是

15. Some unimportant thought runs through my mind and bothers me. / 一些不重要的想法在我脑中反复出现并困扰着我。

1 = Almost Never / 几乎从不 2 = Sometimes / 有时 3 = Often / 经常　4 = Almost Always / 几乎总是

16. I take disappointments so keenly that I can't put them out of my mind. / 我对失望的反应太强烈，无法释怀。

1 = Almost Never / 几乎从不 2 = Sometimes / 有时 3 = Often / 经常　4 = Almost Always / 几乎总是

17. I am a steady person. / 我是一个沉稳的人。

1 = Almost Never / 几乎从不 2 = Sometimes / 有时 3 = Often / 经常　4 = Almost Always / 几乎总是

18. I get in a state of tension or turmoil as I think over my recent concerns and interests. / 一想到我最近关心的事情，我就会感到紧张或烦乱。

1 = Almost Never / 几乎从不 2 = Sometimes / 有时 3 = Often / 经常　4 = Almost Always / 几乎总是

19. I feel satisfied with myself. / 我对自己感到满意。

1 = Almost Never / 几乎从不 2 = Sometimes / 有时 3 = Often / 经常　4 = Almost Always / 几乎总是

20. I feel like I'm losing out on things because I can't make up my mind soon enough. / 因为我无法及时做决定，我觉得自己在错失很多事情。

1 = Almost Never / 几乎从不 2 = Sometimes / 有时 3 = Often / 经常　4 = Almost Always / 几乎总是

## Appendix D

## STAI-State Questionnaire (Form Y-1) 状态焦虑量表（Y-1表）

Instructions 说明：

A number of statements which people have used to describe themselves are given below. Read each statement and then select the response that best describes how you feel right now, at this moment.
下面是一些人们用来描述自己状态的句子。请您仔细阅读每一句话，并选择最能描述您当前（此刻）感受的选项。

Use the following scale 使用下列评分：

1 = Not at all 一点也没有
2 = Somewhat 有一点
3 = Moderately 中等程度
4 = Very much so 非常明显

1. I feel calm. 我感到平静。
    1  2  3  4

2. I feel secure. 我感到有安全感。
    1  2  3  4

3. I am tense. 我感到紧张。
    1  2  3  4

4. I feel strained. 我感到有压力。
    1  2  3  4

5. I feel at ease. 我感到自在。
    1  2  3  4

6. I feel upset. 我感到烦乱。
    1  2  3  4

7. I am presently worrying over possible misfortunes. 我目前在担忧可能的不幸。
    1  2  3  4

8. I feel satisfied. 我感到满足。
    1  2  3  4

9. I feel frightened. 我感到害怕。
       1  2  3  4

10. I feel comfortable. 我感到舒服。
        1  2  3  4

11. I feel self-confident. 我感到自信。
        1  2  3  4

12. I feel nervous. 我感到**紧张**不安。
        1  2  3  4

13. I am jittery. 我感到坐立不安。
        1  2  3  4

14. I feel indecisive. 我感到犹豫不决。
        1  2  3  4

15. I am relaxed. 我感到放松。
        1  2  3  4

16. I feel content. 我感到**满**足安逸。
        1  2  3  4

17. I am worried. 我感到担**忧**。
        1  2  3  4

18. I feel confused. 我感到困惑。
        1  2  3  4

19. I feel steady. 我感到**镇**定。
        1  2  3  4

20. I feel pleasant. 我感到愉快。
        1  2  3  4

**Appendix E**

## NASA Task Load Index (NASA-TLX) Questionnaire

This version of the NASA-TLX is adapted for evaluating subjective workload during hospital navigation tasks (e.g., using AR or paper maps to find a medical destination). Please reflect on your experience during the navigation task when rating each of the following six dimensions.

### Mental Demand

How mentally demanding was the task?

Please mark your perceived level on a scale from 0 to 100.
0 = Very Low | 100 = Very High

### Physical Demand

How physically demanding was the task?

Please mark your perceived level on a scale from 0 to 100.
0 = Very Low | 100 = Very High

### Temporal Demand

How hurried or rushed was the pace of the task?

Please mark your perceived level on a scale from 0 to 100.
0 = Very Low | 100 = Very High

### Performance

How successful were you in accomplishing what you were asked to do?

Please mark your perceived level on a scale from 0 to 100.
0 = Very Low | 100 = Very High

### Effort

How hard did you have to work to accomplish your level of performance?

Please mark your perceived level on a scale from 0 to 100.
0 = Very Low | 100 = Very High

## Frustration

How insecure, discouraged, irritated, stressed, and annoyed were you?

Please mark your perceived level on a scale from 0 to 100.
0 = Very Low | 100 = Very High

Thank you for your participation. Your input is valuable for improving indoor navigation systems in healthcare settings.